\documentclass[9pt,twocolumn]{IEEEtran} 

\def\bg{{\bf g}}
\def\bs{{\bf s}}

\def\by{{\bf y}}

\def\bY{{\bf Y}}

\def\bG{{\bf G}}

\def\b0{{\bf 0}}

\def\bW{{\bf W}}

\usepackage{amsbsy,amsfonts,amsfonts,amssymb,amsmath,color}
\usepackage{graphics}
\usepackage{graphicx}

\usepackage{cite}

\begin{document}
\def\thefootnote{}
%




\title{ A Gibbs Sampling Based MAP Detection Algorithm for OFDM Over Rapidly Varying Mobile Radio Channels\\
}

\author{%
\authorblockN{\large Erdal Panay\i rc\i{\normalsize\authorrefmark{1}}, Hakan Do\u gan{\normalsize\authorrefmark{2}}
and H. Vincent Poor {\normalsize\authorrefmark{3}}}\\

\normalsize\authorrefmark{1} Department of Electronics Engineering,
Kadir Has University, Istanbul, Turkey
\\  \normalsize \authorrefmark{2} Department of Electrical and Electronics Engineering, Istanbul
 University
 \\
 \normalsize\authorrefmark{1} Department of Electrical Engineering,
Princeton University, Princeton, NJ 08544 USA }

\maketitle
\footnotetext{\hspace*{-3ex} E. Panayirci is on  sabbatical leave from Kadir Has University, Istanbul, Turkey.\\
\thanks{This research has been conducted within the NEWCOM++ Network of Excellence in Wireless Communications and WIMAGIC Strep projects funded through
the EC 7th Framework Programs and was supported in part by the U.S. National Science Foundation under Grant CNS-0625637.}}
\thispagestyle{empty}
\begin{abstract}
In orthogonal frequency-division multiplexing (OFDM) systems
operating over  rapidly time-varying  channels, the orthogonality
between subcarriers is destroyed leading to inter-carrier
interference (ICI) and resulting in an irreducible error floor. In
this paper, a new and low-complexity  maximum {\em a
posteriori} probability (MAP) detection  algorithm is proposed for OFDM systems operating
 over rapidly  time-varying multipath channels.  The detection algorithm exploits
the banded structure of the frequency-domain channel matrix whose
bandwidth is a parameter to be adjusted according to the speed of
the mobile terminal. Based on this assumption, the received signal
vector is decomposed into reduced dimensional sub-observations in
such a way that all components of the observation vector
contributing to the symbol to be detected are included in the
decomposed observation model. The data symbols are then detected by
the MAP algorithm  by means of a Markov chain Monte Carlo (MCMC)
technique in an optimal and computationally efficient way.
Computational complexity investigation as well as simulation results
indicate that this algorithm has significant performance and
complexity advantages over  existing suboptimal detection and
equalization algorithms
proposed earlier in the literature.\\

\end{abstract}



%


\section{Introduction}
Orthogonal frequency-division multiplexing (OFDM) has been shown to be an effective method to overcome
inter-symbol interference (ISI) caused by frequency-selective
fading with a simple transceiver structure, and is consequently  used in several existing wireless local and metropolitan area standards such as the IEEE 802.11 and IEEE 802.16 families. IEEE 802.11 wireless LAN
(WLAN) technology has become very popular for providing  data services
to Internet users although its overall design and feature set are not
well suited for outdoor broadband wireless access (BWA)
applications\cite{ni2007ibr}. Therefore,  IEEE 802.16 has been
developed as a new standard for BWA applications
\cite{eklund2002ist}. Recently, the much-anticipated
Worldwide Interoperability for Microwave Access  (WiMAX) technology was introduced to promote the 802.16 standards while introducing  features
to enable mobile broadband services at  vehicular speeds beyond
120 km/h.

OFDM eliminates ISI and simply uses a one-tap equalizer to compensate multiplicative channel distortion in
quasi-static channels. However, in fading channels with very high mobility, the time
variation of the channel over an OFDM symbol period results in
a loss of subchannel orthogonality which leads to inter-carrier
interference (ICI). A considerable amount of research
on OFDM receivers for  quasi-static fading  has been
conducted, but a major hindrance to such receivers is the lack of mobility
support \cite{hakan07}. Since mobility support is
widely considered to be one of the key features in wireless communication
systems, and in this case ICI degrades the performance of OFDM
systems, OFDM transmission over very rapidly
time varying multipath fading channels has been considered recently in a number of recent works
\cite{leus5,Stamoulis6,huang7,Jeon8,ruguni9,ruguni11,schniter10,Gorokhov12,Choi13,Cai14,Tomasin15,Hwang17}.

The techniques proposed in these works  range from linear
equalizers, based on the zero-forcing (ZF) or the minimum
mean-squared error (MMSE) criterion
\cite{leus5,Stamoulis6,huang7,Jeon8,ruguni9,schniter10,ruguni11,Gorokhov12,Choi13,Cai14},
to nonlinear equalizers based on decision-feedback or ICI
cancelation
\cite{schniter10,ruguni11,Gorokhov12,Choi13,Cai14,Tomasin15}. Also
near maximum-likelihood approaches have been proposed \cite{Ohno}.
It has been shown that nonlinear equalizers based on ICI cancelation
generally outperform linear approaches
\cite{Choi13,Cai14,schniter10,Gorokhov12}. However, linear
equalizers still preserve their importance mainly because they are
less complex.

In \cite{Choi13}, performances of Matched Filter (MF), Least Squares
(LS), Minimum Mean Square Error (MMSE) and MMSE with Successive
Detection (SD) techniques with optimal ordering have been investigated.
However, since the number of subcarriers is usually very large in
high speed wide-band wireless standards, even the linear MMSE equalizer considered
in  \cite{Choi13} demands very high computational load.

The specific structure of the Doppler-induced ICI in OFDM systems
operating over highly mobile channels presents a distinctive feature
of limited support of the Doppler spread that can be exploited by
the receiver.  References \cite{huang7,
Jeon8,ruguni9,schniter10,ruguni11} exploit the
banded character of the frequency-domain channel matrix to reach a
complexity that is only linear in the number of subcarriers.  In a
certain sense, the assumption of a banded frequency-domain channel
matrix is a natural extension of the time-invariant channel case,
in which the frequency-domain channel matrix is diagonal and hence
banded with the smallest possible bandwidth.

In \cite{Jeon8}, using the banded structure of the channel matrix, a simple
frequency domain equalizer has been proposed that can compensate for the loss of subchannel
orthogonality due to ICI. However, the detection performance of the technique
degrades substantially, since  the data to be detected cannot fully
use the contributing observation elements. The work presented in
\cite{Cai14} combined \cite{Jeon8} and \cite{schniter10} to derive
 a recursive decision feedback equalizer receiver for ICI suppression.
The iterative MMSE serial linear equalizer (SLE) of
\cite{schniter10}, which takes the banded structure of the channel matrix into
account, seems to be one of the most promising approaches to
compensate for ICI. Iterative MMSE is then applied to estimate
frequency-domain symbols. In \cite{ruguni9}, a block MMSE equalizer
for OFDM systems operating over time-varying channels is presented. By
exploiting the banded structure of the frequency-domain channel matrix, the
complexity of the resulting algorithm turns out to be smaller than
that of \cite{schniter10}.

In this paper, a new computationally feasible, maximum {\em a
posteriori} probability (MAP)-based  data symbol detection algorithm is proposed
for OFDM systems operating over highly mobile channels, as an
alternative to the existing suboptimal equalization/detection
techniques summarized in the above paragraphs. The proposed
detection algorithm exploits the banded structure of the
frequency-domain channel matrix whose bandwidth is a parameter to be
adjusted according to the  speed of the mobile terminal. This
assumption enables us to decompose the main received signal
vector into finite numbers of  reduced-dimensional, sub-received
signal vectors from which the data symbols can be detected by the
MAP algorithm in an optimal and computationally efficient way. The
decomposition is achieved in such a way that  all the components of
the received vector that contribute to the symbol to be detected are
included in the decomposed observation model. Data symbols in each
sub-received signal model are then detected successively by  a MAP
detection algorithm. To implement  MAP symbol detection in a
computationally efficient way, we employ a Markov Chain Monte Carlo
(MCMC) technique based on Gibbs sampling, which is a powerful
statistical signal processing tool to estimate {\em a posteriori}
 probability (APP) values.

The resulting detection algorithm is compared with previously
proposed algorithms in terms of both bit error rate (BER) and
complexity requirements. Computational complexity investigation as
well as simulation results indicate that our algorithm has
significant performance and complexity advantages over the existing
suboptimal detection and equalization algorithms.

\section{System Model}
Let us consider an OFDM system with $N$ subcarriers and  available
bandwidth $B=1/T_s$ where $T_s$ is the sampling period. A given
sampling period is divided into $N$ subchannels by equal frequency
spacing $\Delta f=B/N$. At the transmitter, information symbols are
mapped into possibly complex-valued transmitted symbols according to
the modulation format employed. The symbols are processed by an
$N-$length Inverse Fast Fourier Transform (IFFT) block that
transforms the data symbol sequence into the time domain. The
time-domain signal is extended by a guard interval containing  $G$
samples whose length is chosen to be longer than the expected delay
spread to avoid ISI. The guard interval includes a cyclically
extended part of the OFDM block to avoid ICI. Hence, the complete OFDM
block duration is $P=N+G$ samples. The resulting signal is converted
to an analog signal by a digital-to-analog (D/A) converter. After
shaping with a low-pass filter (e.g. a raised-cosine filter) with
bandwidth $B$, it is transmitted through the transmit antenna with
the overall symbol duration of $PT_{s}$.

Let $h(m,l)$ represent the $l$th path (multipath component) of the
time-varying channel impulse response at time instant $t=m T_s$. The
discrete-time received signal can then be expressed as follows:
\begin{equation}\label{2}
    y^{}(m)=\sum_{l=0}^{L-1}h^{}(m,l)d^{}(m-l)+w(m),
\end{equation}
where the transmitted  signal $d(m)$ at discrete sampling time $m$ is
given by
\begin{equation}\label{1}
d^{}(m)=\frac{1}{\sqrt{N}}\sum_{k=0}^{N-1}s^{}(k) e^{j2\pi mk /N},
\end{equation}
$L$ is the total number of paths of the frequency selective fading
channel, and $w(m)$ is  additive white Gaussian noise (AWGN) with
zero mean and variance $E\{|w(m)|^2\}=\sigma_w^2$. The sequence
$s(k),k=0,1,\cdots, N-1,$ in (2) represent either
quadrature-amplitude modulation (QAM) or phase-sift-keying (PSK)
modulated data symbols with $E\{|s(k)|^{2}\}=1.$

At the receiver, after passing through the analog-to-digital (A/D)
converter and removing the cyclic prefix (CP), a fast Fourier
transform (FFT)  is used to transform the data back into the
frequency domain. Lastly, the binary data is obtained after
demodulation and channel decoding.

The fading channel coefficients $h(m,l)$ can be  modeled as
zero-mean complex Gaussian random variables. Based on the wide-sense
stationary uncorrelated scattering (WSSUS) assumption, the fading
channel coefficients in different paths are uncorrelated with each
other. However, these coefficients are correlated  within each
individual path and have a Jakes Doppler power spectral density
 having an autocorrelation function given by
\begin{equation}\label{chanxcorr}
    E\{h^{}(m,l)h^{*}(n,l)\}=\sigma^2_{h_{l}}J_0{(2\pi
    f^{}_dT_s(m-n))},
\end{equation}
where $\sigma^2_{h_{l}}$ denotes the power of  the channel
coefficients of the $l$th path.  $f^{}_d$ is the Doppler frequency
 in Hertz so that the  term $f_{d}T_{s}$ represents the
normalized Doppler frequency of the channel coefficients. $J_0(.)$
is the zeroth order Bessel function of the first kind.

By using (\ref{1}) in (\ref{2}), the received signal can be written as
\begin{equation}
\label{recsigsingl1}
y(m)=\frac{1}{\sqrt{N}}\sum_{k=0}^{N-1}s(k)\sum_{l=0}^{L-1}h(m,l)e^{j\frac{2\pi
k(m-l)}{N}}+w(m),
\end{equation}
which upon defining  the time-varying channel transfer function
\begin{equation}
H(k,m)\triangleq {\sum}^{L-1}_{l=0}h(m,l)e^{-j2\pi lk/N},
\end{equation}
 becomes
\begin{equation}
\label{recsigsingl2}
y(m)=\frac{1}{\sqrt{N}}\sum_{k=0}^{N-1}s(k)H(k,m)e^{j2\pi mk /N}+w(m).
\end{equation}
The FFT output at the $k^{th}$ subcarrier, after excluding the guard interval, can be expressed as
\begin{eqnarray}\label{FreqDomRecSig}
\nonumber    Y(k)&=&\frac{1}{\sqrt{N}}\sum_{m=0}^{N-1}y(m)e^{-j2\pi
mk /N}\\
 &=&s(k)G(k,k)+I(k)+W(k),
\end{eqnarray}
where
$I(k)$ is ICI caused by the time-varying nature of the channel given
as
\begin{equation}\label{ICI}
 I(k)=\sum_{i=0 , i \neq k
}^{N-1}s(i)G(k,i).
\end{equation}

$G(k,i)$ in (\ref{ICI}) represents the average frequency domain time-varying channel response, defined as
\begin{equation}\label{ICI2}
G(k,i)\triangleq (1/N) \sum_{m=0}^{N-1}H(i,m)e^{j2\pi m(i-k) /N}.
\end{equation}

Similarly, the term $G(k,k) \triangleq\frac{1}{N}\sum_{m=0}^{N-1}H(k,m)$ in (\ref{ICI2}) represent the portion of the average frequency domain channel response at the $k$th subcarrier and $W(k)$  denotes discrete Fourier transform of the white Gaussian noise $w(m)$:
\begin{equation}\label{FreqDomNoise}
    W(k)=\frac{1}{\sqrt{N}}\sum_{m=0}^{N-1}w(m)e^{-j2\pi mk /N}.
\end{equation}

Because of the term $I(k)$ in (\ref{FreqDomRecSig}), there is an irreducible error
floor even in the training sequences since pilot symbols are also
corrupted by ICI, arising from the fact that the time-varying channel
destroys the orthogonality between subcarriers. Therefore, channel
estimation should be performed either jointly with data or before the FFT block in order to compensate for the ICI.
Note that if the channel is static or very slowly time-varying, that is $H(k,m)\thickapprox H(k)$, then it can be easily shown
that $G(k,k)=H(k)$ and $I(k)=0$ for $k=0,1,\cdots,N-1,$ resulting in a received signal at the output of the FFT processor
corresponding to  the  $k$th OFDM symbol given by
\[ Y(k)=H(k)s(k)+W(k).\]

From (6) and (7), the FFT output received signal  can be expressed in  vector form as
\begin{equation}\label{genel}
 \bf{Y}=\bf{G} \bf{s}+\bf{W}
\end{equation}
where ${\bf{Y}}=[Y(0),Y(1),...,Y(N-1)]^{T}$, ${\bf{s}}=[s(0),s(1),...,s(N-1)]^{T}$ and ${\bf{W}}=[W(0),W(1),...,W(N-1)]^{T}.$
 For $k,i=0,1,\cdots,N-1,$ the $(k,i)$th element of the matrix
 $\bG=[G(k,i)]\in \mathcal{C}^{N\times N}$ representing the time-varying channel  is given by
 (9).

Under the assumption that the channel matrix $\bG$ in (11) is
perfectly known at the receiver, the maximum likelihood (ML)
detector performs an exhaustive search over the entire set of signal
vectors whose components are selected from the signal constellation
formed by the modulation scheme chosen. Especially in IEEE 802.16
based systems, the length $N$ of each OFDM symbol is very large; it
can take values as large as  $N=1024$ or even $N=2048$ especially
for high mobility applications. In this case an exhaustive search
for the ML solution would be very complex since the search space has
an extremely large  number of constellation points,
($|\mathcal{S}|^{N}$, where $|\mathcal{S}|$ is the cardinality of
the signal constellation).  On the other hand, all of the lower
complexity linear detectors given in Table 1 are suboptimal
since they do not take into account the correlation of the
components of the transformed noise and  yield noise enhancement.
Recently, a nonlinear recursive detection technique using the
decision-feedback principle, namely the  MMSE-SD algorithm (VBLAST),
has been proposed \cite{Choi13}.  The performance of VBLAST depends
critically on the order in which the data vector components are
processed. To minimize error propagation effects and to improve the
detection of unreliable components, more reliable data vector
components should be detected first. Therefore, the algorithm
depends on calculation of the post-detection signal-to-interference-plus-noise
 ratio (SINR) based upon MMSE detection as a measure  of
reliability, and so the calculation of SINR is compulsory at each iteration.
Therefore this algorithm is computationally intensive as a number of pseudo inverse operations need
to be performed. Moreover, its complexity  grows exponentially with
the total number of subcarriers.
\begin{table}[h]
  \centering
\caption{Linear Detection Methods}\label{y2}
\begin{tabular}{|l|l|l|}
  \hline
  Method & Solution \\  \hline
  Matched Filter (MF) & $\bf{\hat{s}}=\mbox{Q} \{\bf{G^{\dagger}} \bf{Y}\}$ \\  \hline
  Zero Forcing (ZF) & $ \bf{\hat{s}}= \mbox{Q} \{(\bf{G^{\dagger}} \bf{G})^{-1}  \bf{G^{\dagger}} \bf{Y}\}$ \\  \hline
  MMSE & $\bf{\hat{s}}= \mbox{Q} \{(\bf{G^{\dagger}} \bf{G}+{\bf I}_N \sigma_w^2)^{-1}  \bf{G^{\dagger}} \bf{Y}\}$
  \\ \hline
\end{tabular}
\end{table}

where ${\bf I}_N$ is the $N$-by-$N$ identity matrix.
\section{MAP Detection  Algorithm }
The necessity of detecting  large numbers of symbols  in OFDM  systems employed especially in highly mobile and wide-band wireless systems represents a significant computational burden as well as creating some convergence problems. However, as is known \cite{ruguni11}, time-varying channels produce
 a nearly-banded channel matrix $\bG$ whose only  main diagonal,  $Q$ subdiagonals and  $Q$ superdiagonals are nonzero. The bandwidth $2Q+1$, which is
 defined as the total number of non-zero diagonals in $\bG$, is  a parameter to be adjusted according to the mobility-rate of the channel.
 The significant coefficients of $\bG$ thus are confined to the $2Q+1$ central diagonals. The parameter $Q\in \{0,1,\cdots N/2-1\}$ controls the
 target ICI response length; larger $Q$ corresponds to a longer ICI span and, thus, increased estimation complexity. In general, $Q$ should be chosen proportional to the width of the Doppler spectrum of the channel.

 We now present an optimal low-complexity MAP detection algorithm to detect the data  symbols $\bs$ from $\bY$ taking into account
 the banded structure of the channel matrix $\bG$.
 From the observation $\bY$ in (11), the receiver attempt to detect the OFDM output symbol $\bs$, assuming that $\bG$ is completely
 known by the receiver.  The banded structure of the channel
 matrix $\bG$ implies that the data symbol $s(k), k=0,1,\cdots, N-1,$ contributes to a maximum of $2Q+1$ observation elements
 as follows:
 \begin{equation}
 \bY_{k}=[Y(j_{k}),  Y(j_{k}+1),\cdots, Y(i_{k})]^{T}, \mbox{   for}
 \end{equation}
 \[ j_{k}=\max\{0,k-Q\} \hspace{3mm} \mbox{and} \hspace{3mm} i_{k}=\min\{N-1,k+Q\}.\]

 Based on this observation, the received signal in (11) can be decomposed into $N$ reduced-dimensional
 sub-observations from which the data symbols can be detected in an optimal  and  computationally efficient way.
 For a given index $k=0,1,\cdots, N-1$  and $Q$,  it can be easily shown from (11) and (12) that

 \begin{equation}
 \bY_{k}=\bG_{k}\bs_{k}+\bW_{k}
 \end{equation}
 where,
 $\bs_{k}=\left[\bs(j_{k})\right]$,
 $\bW_{k}=\left[\bW(j_{k})\right]$,
 $\bG_{k}=\left[ G(i_{k},j_{k})\right]$, for
 \begin{equation}
I_{L_{k}}\triangleq\max\{0,k-Q\}\leq i_{k}\leq I_{U_{k}}\triangleq\min\{N-1,k+Q\}
\end{equation}
and
\begin{equation}
J_{L_{k}}\triangleq\max\{0,k-2Q\}\leq j_{k}\leq
J_{U_{k}}\triangleq\min\{N-1,k+2Q\}.
 \end{equation}

 Note that due to the banded structure  of $\bG$, some elements of the matrices $\bG_{k}$ are
 zero and $\mbox{dim}(\bG_{k})\leq (2Q+1)\times (2(2Q+1)-1)$. The $\bG_{k}$'s reach their maximum dimension when
 $2Q+1\leq k \leq (N-(2Q+1))$.

 For $k=0,1,\cdots, N-1,$ the MAP estimate of the data symbol
 $s(k)$ given $\bY_{k}$ is
 \begin{equation}
 \widehat{s}_{MAP}(k)\equiv\widehat{s}(k)=\arg\max_{s(k)\in \mathcal{S}} P(s(k)|\bY_{k}),
 \end{equation}
where $\mathcal{S}$ denotes the set of signal constellation points from which $s(k)$ takes values. Based on this approach, $s(k)$ can be detected
sequentially for $k=0,1,\cdots,N-1$, incorporating the outcomes of the
 previous estimates in a decision-feedback mode as follows.\\

$ \blacksquare $ For $k=0$, determine the estimate  $\widehat{s}(0)$
from (16).

 $ \blacksquare $ For $k=k+1 $ modify the observation vector $\bY_{k}$ by subtracting the terms
 coming from the contributions of the estimated data symbols $\widehat{s}(0),\widehat{s}(1),\cdots, \widehat{s}(k-1)$ as
 \begin{equation}
 \widetilde{\bY}_{k}\triangleq\bY_{k}-\sum_{l=J_{L_{k}}}^{k-1}\bg^{(l)}_{k}\widehat{s}(l)
 \end{equation}
 where $\bg^{(l)}_{k}$ is the $l$th column of $\bG_{k}$ and $J_{L_{k}}$ and $J_{U_{k}}$
 are defined in (15).

 $ \blacksquare $ Determine the MAP estimate of $\widehat{s}(k)$ from
 \begin{equation}  \widetilde{\bY}_{k}=\widetilde{\bG}_{k}\widetilde{\bs}_{k}+\bW_{k}\end{equation}
 as
 \begin{equation} \widehat{s}(k)=\arg\max_{\widetilde s(k)\in \mathcal{S}}P(\widetilde s(k)|\widetilde{\bY}_{k}),\end{equation}
 where
 \begin{equation}
 \widetilde{\bG}_{k}\triangleq \bG_{k}-\left[\bg^{(0)}_{k}, \bg^{(1)}_{k},\cdots,\bg^{(k-1)}_{k}\right],
 \end{equation}
 and $\widetilde{\bs}_{k}$ is the vector obtained by removing the first $k-1$ elements of
 $\bs_{k}.$\\
$ \blacksquare $ END IF $k=N-1$.

The major problem is finding the values of $\hat{s}_{MAP}(k)$ in a computationally efficient manner.
To see this difficulty we assume that the data symbols are independent and identically distributed binary phase shift
keying (BPSK), taking values of $+1$ and $-1$. Note that higher dimensional signal constellations can be treated similarly
with a straightforward extension. The conditional probability of $\widetilde s(k)$ given the observation vector $\bf \widetilde Y_{k}$ can be expressed as

\begin{eqnarray}
\hspace{-8mm}P( \widetilde s(k)= +1|\widetilde{\bY}_{k})&\hspace{-2mm}=&\hspace{-3mm}\sum_{\widetilde{\bs}_{\overline{k}}}P(\widetilde s(k)=+1,\widetilde{\bs}_{\overline{k}}|\widetilde{\bY}_{k}) \nonumber\\
                &\hspace{-6mm}=&\hspace{-5mm}\sum_{\widetilde{\bs}_{\overline{k}}}  P(\widetilde s(k)= +1|\widetilde{\bs}_{\overline{k}},\widetilde{\bY}_{k})P(\widetilde{\bs}_{\overline{k}}|\widetilde{\bY}_{k})
\end{eqnarray}
where the second identity follows by applying the chain rule of probability.
The vector  $\bf \widetilde s_{\overline{k}}$ in (21) is obtained by canceling the component
$\widetilde s(k)$ in $\bf \widetilde s_{k}$ and the summation   is over all possible values of $\bf \widetilde s_{\overline{k}}$.
The number of combinations that $\bf \widetilde s_{\overline{k}}$ takes grows exponentially with the dimension of
$\bf \widetilde s_{k}$ and thus becomes prohibitive for large values of the size of this vector. Thus, we resort to the Gibbs sampler,
a Monte Carlo method to calculate the {\em a posteriori} probabilities of the unknown symbols.

\subsection{MAP Detection Based on Gibbs Sampling}
The Gibbs sampler is an MCMC sampling
method for numerical evaluation of multidimensional integrals.
 Its popularity is gained from the facts that it is capable of
carrying out many complex Bayesian computations. In this section we
briefly explain the application of the Gibbs sampling technique to our
symbol detection problem where the observation process is given by
(18). For notational convenience we drop the index $k$ and the
"tilde"  from all the involved variables, e.g.
$\bY,\bG$ are shorthand notations for $\widetilde{\bY}_{k},\widetilde{\bG}_{k}$, respectively. (21) can then
be expressed as

\begin{eqnarray}
P(s(k)= +1|\bY) &=&\sum_{\bs_{\overline{k}}} P(s(k)= +1|\bs_{\overline{k}},\bY)P(\bs_{\overline{k}}|\bY)\nonumber\\
                &=& E_{\bs_{\overline{k}}|\bY}\left\{P(s(k)= +1|\bs_{\overline{k}},\bY)\right\}.
\end{eqnarray}
According to the Gibbs sampling based statistical Monte Carlo estimation technique, an estimate of (22) can be evaluated
by taking the empirical average
\begin{equation}
P(s(k)= +1|\bY)=\frac{1}{N_s}\sum_{n=1}^{N_{s}}P(s(k)= +1|\bs^{(n)}_{\overline{k}},\bY)
\end{equation}
where $\bs^{(n)}_{\overline{k}}$ for  $n=1,2,\cdots, N_{s}$ are samples drawn from the conditional
distribution $P(\bs_{\overline{k}}|\bY)$. There is a substantial body of literature concerning the Monte Carlo Gibbs
sampling technique; see, e.g., \cite{behrouz},\cite{casella}. One possible version  of the Gibbs sampler suitable for
calculating the {\em a posteriori} probabilities in (21) may be summarized as follows.

Let $\bs=[s(0),s(1),\cdots,s(N-1)]^{T}$ be a vector of unknown data symbols. Let $\bY$ be the observed signal.
To generate random samples from the distribution $P(\bs|\bY)$, given the samples from the
$(n-1)$th iteration $\bs^{(n-1)}=[s^{(n-1)}(0),s^{(n-1)}(1),\cdots,s^{(n-1)}(N-1)]^{T}$,
the Gibbs  algorithm iterates at the $n$th iteration as follows to generate the samples
$\bs^{(n)}=[s^{(n)}(0),s^{(n)}(1),\cdots,s^{(n)}(N-1)]^{T}$:\\
$\blacksquare$ Initialize $\bs^{(0)}$ randomly;\\
$\blacksquare$ for  $n=1,2,\cdots,N_{T}$ and for $k=0,1\cdots,N-1,$\\
draw sample $s^{(n)}_{k}$
 from $ P\left(s(k)|s^{(n)}_{0},\cdots,s^{(n)}_{k-1},s^{(n-1)}_{k+1},\cdots,s^{(n)}_{N-1}\right).$

 Note that to ensure convergence, the Gibbs iteration is usually carried out for  $N_{T}=N_b+N_s$ iterations.
 The first $N_{b}$ iterations of the loop is called the {\em burn-in}
 period which is necessary for the Monte Carlo simulation to  reach its  stationary
 distribution. Only the samples $\bs^{(n)}=[s^{(n)}_{0},s^{(n)}_{1},\cdots,s^{(n)}_{N-1}]^{T}, n=N_{b}+1,\cdots, N_{T}$, from the last
 $N_{s}$ iterations are used to calculate the expectation.

 It is known that under regularity conditions \cite{geman,casella},

 \noindent (i) the distribution of $\bs^{(n)}$ converges to $P(\bs|\by)$, as $n\rightarrow \infty$.   \\
 \noindent (ii) $(1/N_{s}\sum_{n=1}^{N_{s}}P(s(k)= +1|\bs^{(n)}_{\overline{k}},\bY)=
 \sum_{\bs_{\overline{k}}} P(s(k)= +1|\bs_{\overline{k}},\bY)P(\bs_{\overline{k}}|\bY)$, as $n\rightarrow \infty$.

 \subsection{Implementation of  the Symbol Detector}

 From the previous section, we recall that to compute
 $P(s(k)= +1|\bY_{k})=\sum_{\bs_{\overline{k}}}P(s(k)=+1,\bs_{\overline{k}}|\bY_{k}) $, we need to perform the summation on the right-hand
 side of (22). When $\bs$ has a large dimension, the exact evaluation of this summation may not be feasible and other more efficient
 techniques must be adopted. In this section the Gibbs sampling-based Monte Carlo method summarized in the previous section will be applied
 to develop a computationally efficient algorithm for calculation of the {\em a posteriori} probabilities $P(s(k)|\bY).$
 From (23), it follows that we need to evaluate $P(s(k)= +1|\bs^{(n)}_{\overline{k}},\bY),$ for $n=1,2,\cdots,N_{T}$. For this we define
 \begin{equation}
 \lambda^{(n)}_{k}\triangleq\ln \frac{P(s(k)= +1|\bs^{(n)}_{\overline{k}},\bY)}{ P(s(k)= -1|\bs^{(n)}_{\overline{k}},\bY)},
 \end{equation}
from which it can be easily seen that
\begin{equation}
P(s(k)= +1|\bs^{(n)}_{\overline{k}},\bY)=\frac{1}{1+\exp\left(-\lambda^{(n)}_{k}\right)}.
\end{equation}

 $\lambda^{(n)}_{k}$ can be computed by expanding $P(s(k)= +1|\bs^{(n)}_{\overline{k}},\bY)$ as

 \begin{eqnarray} &&P\left(s(k)= +1|\bs^{(n)}_{\overline{k}},\bY\right)=\\
 &&\frac{ p\left(\bY|s(k)=+1,\bs^{(n)}_{\overline{k}})P(s(k)=+1,\bs^{(n)}_{\overline{k}}\right)}
 {\sum_{s(k)\in\{+1,-1\}}p\left(\bY|s(k)=+1,\bs^{(n)}_{\overline{k}}\right)P\left(s(k)=+1,\bs^{(n)}_{\overline{k}}\right)}.\nonumber
 \end{eqnarray}

The data symbols are assumed to be independent and equally likely, Therefore, it follows from (26) and (24)
that
 \begin{equation}
 \lambda^{(n)}_{k}\triangleq\ln \frac{p( \bY| s(k)= +1, \bs^{(n)}_{\overline{k}})}{ p(\bY| s(k)= -1, \bs^{(n)}_{\overline{k}})}.
 \end{equation}
Since $p(\bY|\bs)\thicksim \exp(-|\by-\bG \bs|^{2})$, after some algebra, (27) can be expressed as
\begin{equation}
 \lambda^{n}(s_{k})=\frac{1}{\sigma^{2}_{w}} \Re\left\{\bg^{\dag}_{k}(\bY-\bG_{\overline{k}} \mbox{  } \bs_{\overline{k}})\right\},
 \end{equation}
where $(\cdot)^{\dag}$ denotes the conjugate transpose and $\Re\{\cdot\}$ denotes  the real part of its argument. $\bG_{\overline{k}}$ is
$\bG$ with its $k$th column $\bf{g}_{k}$
removed. In summary, for $k=0,1,\cdots,N-1$, to estimate the {\em a posteriori} probabilities $P(s(k)|\bY)$ in (21), the Gibbs sampler runs over
all symbols $N_{s}$ times to generate a collection of vectors $\left\{\bs^{(n)}_{\overline{k}}\right\}_{n=N_{b}+1}^{N_{T}}$ which are used in (23) to compute the desired quantities.
\subsection{Complexity Requirements} The computational  complexity of the MAP symbol detector based on Gibbs
sampling proposed in this work  is determined by the parameters
$N_{s}, Q, N$ and the constellation size of the transmitted data
symbols. The computation of $\widetilde{\bY}_{k}$ in (18) for
$k=0,1,\cdots,N-1$ requires a maximum of  $(4Q^{2}+2Q)N$ complex
multiplications (CMs) and $(4Q^{2}+2Q)N$ complex additions (CAs) per data
block. Assuming BPSK signaling, the computation of the {\em a
posteriori} probabilities in (22) requires a maximum of
$(8Q^{2}+2Q+1)NN_{s}$ CMs and $(8Q^{2}+2Q-1)NN_{s}$ CAs and
computation of the empirical average of {\em a posteriori}
probabilities of the data symbols in (23) requires $NN_{s}$ CSs.
Therefore, the whole algorithm requires of maximum
$N(4Q^{2}+2Q+(4Q^{2}+2Q)N_{s})$ CMs and
$N(4Q^{2}+2Q+(4Q^{2}+2Q)N_{s})$ CAs, leading to a total of
$2N(4Q^{2}+2Q+(4Q^{2}+2Q)N_{s})$ complex operations.

Several low complexity equalization algorithms have been developed recently, of which
several are worth  mentioning here to compare their computational complexities with that of the Gibbs-based algorithm.

Ruguni et al. \cite{ruguni9}  proposed a block MMSE technique based on  exploiting the banded structure
of the channel matrix $\bG$. The matrix inversion was obtained using a low-complexity decomposition such as
Cholesky or the $LDL^{\dag}$ decomposition. The algorithm requires a total of $(8Q^{2}+22Q+4)N$ complex operations.
Schniter \cite{schniter10} proposed  a linear serial equalizer also based on  exploiting the banded structure of the channel matrix.
This algorithm requires a total of $(8/3Q^{3}+2Q^{2}+5/3Q+4)N$ complex operations. The complexity of the serial MMSE equalizer is higher than
that of the block MMSE equalizer.

In the VBLAST algorithm, matrix inversion is needed of dimension equal to
the number of OFDM subcarriers. As a result, the computational
complexity of the VBLAST receiver increases rapidly with the
number of subcarriers, which makes its real-time implementation
prohibitive for large numbers of subcarriers.

As can be seen easily, the complexity of our algorithm  is of the same order of the above equalization algorithms and is
lower  than the VBLAST algorithm. However, as remarked earlier, these algorithms are suboptimal  as opposed to our optimal MAP detection algorithm
and perform poorly especially when the ICI is high.
It is also worth mentioning that our algorithm can be easily extended to an iterative multiuser MAP detection scheme for OFDM systems.

\section{Simulation Results}
This section presents computer simulation results of the proposed
detection methods for rapidly varying mobile radio channels.  The
system operates with a 5 MHz bandwidth and is divided into 512 tones
($N=512$) with a total symbol period ($T_s$) of 115 $\mu$s, of which
12.8 $\mu$s constitute the CP. One OFDM symbol thus consists of 576
samples, sixty-four of which constitute the CP. The normalized
Doppler frequencies are $f_{d1}*T_s=0.0307$ and $f_{d2}*T_s=0.1075$,
corresponding to an IEEE 802.16e mobile terminal moving with speeds
of 120 km/h and 420 km/h, respectively, for a carrier frequency of
2.4 GHz. The wireless channels between the mobile antenna and the receiver
antenna are modeled based on a realistic channel model determined by
the COST-207 project in which Typical Urban (TU)  and Bad Urban (BU) channel models are considered. For each OFDM symbol,  Gibbs sampling is performed for 30 iterations, with the first 10 iterations as the "burn-in" period.

\begin{figure}
\includegraphics[width=90mm,height=65mm]{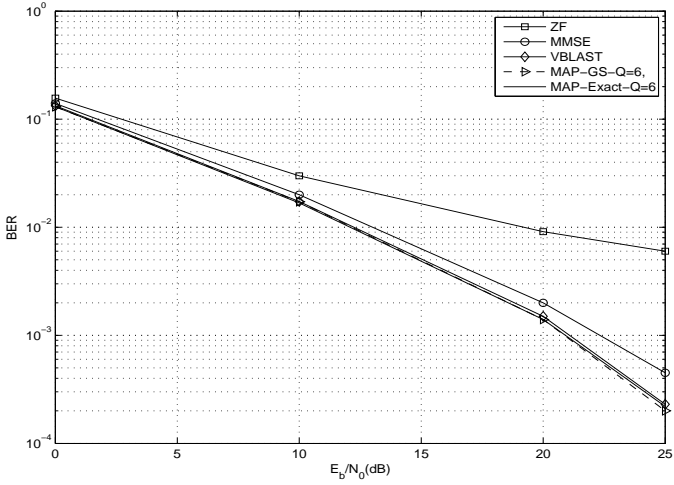}
\caption{BER comparison of various detection algorithms for OFDM systems; TU Channel (6 taps), 420 km/h}\label{fig:rec1}
\end{figure}

\begin{figure}
\includegraphics[width=90mm,height=65mm]{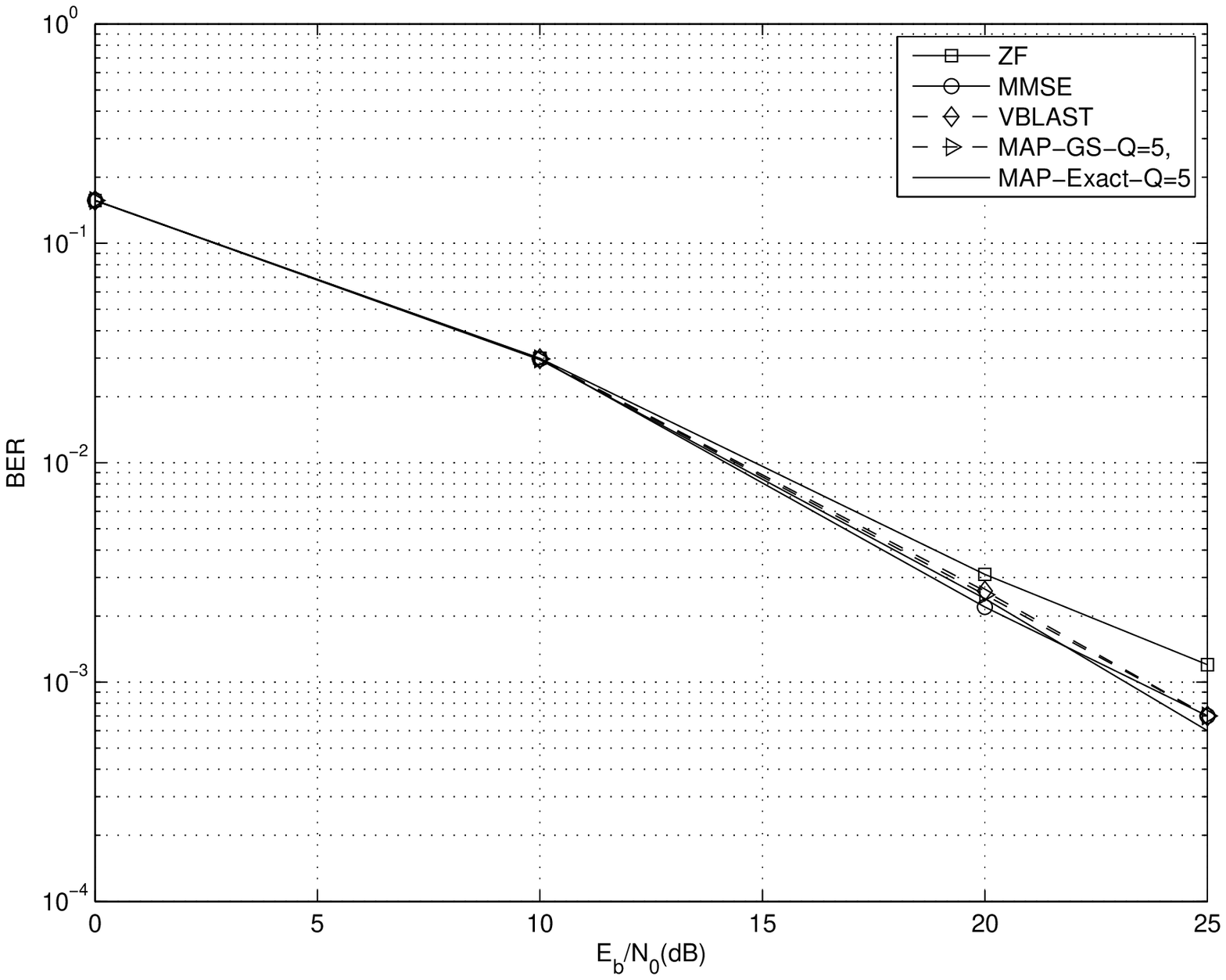}
\caption{BER comparison of various detection algorithms for OFDM systems; TU Channel (6 taps), 120 km/h}\label{fig:rec2}
\end{figure}

\begin{figure}
\includegraphics[width=90mm,height=65mm]{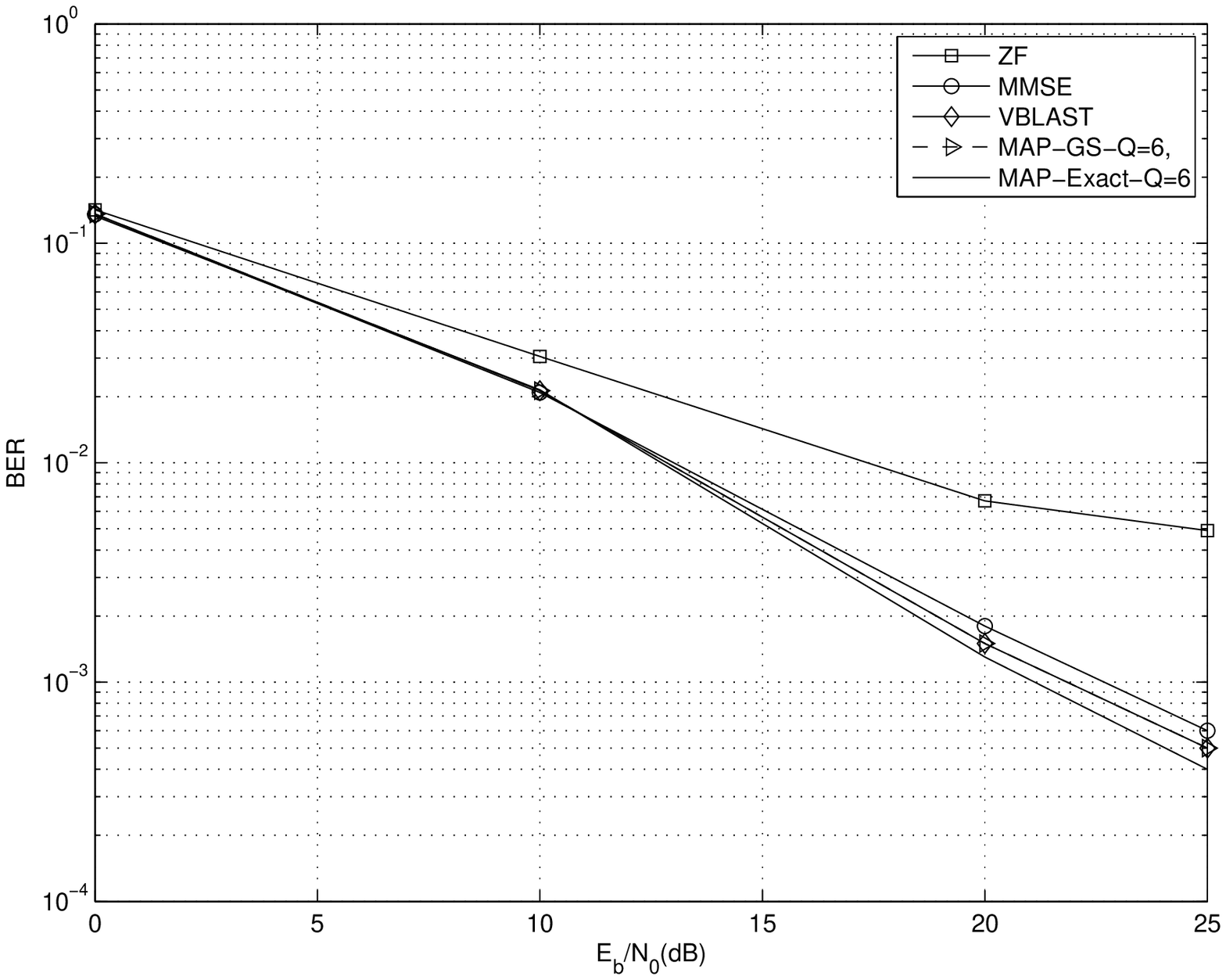}
\caption{BER comparison of various detection algorithms for OFDM systems; BU Channel (6 taps), 420 km/h}\label{fig:rec3}
\end{figure}

\begin{figure}
\includegraphics[width=90mm,height=65mm]{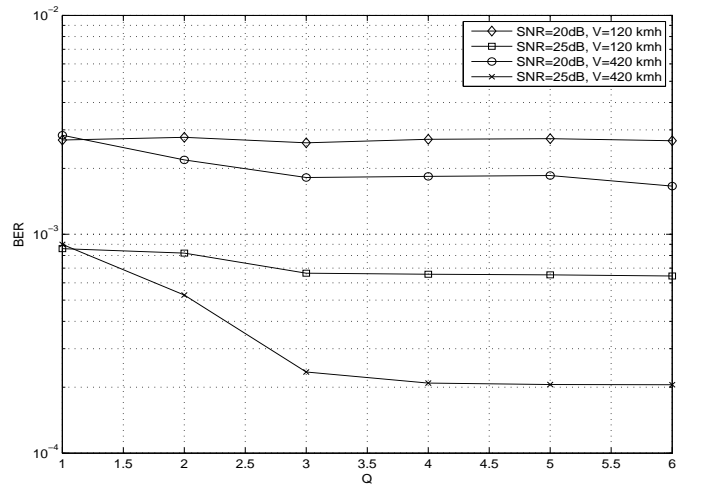}
\caption{BER performance of the proposed MAP algorithm based on Gibbs Sampling for the  TU channel as a function of  Q values}\label{fig:rec4}
\end{figure}

%
%
%

The information symbols are BPSK modulated to yield $\mathcal{S}$ =
$\{ \pm 1 \}$. Figs.1-3 compare  the BER performance of the proposed
Gibbs-based MAP detection algorithm, an equalization technique based
on zero forcing (ZF) proposed in \cite{Jeon8}, linear MMSE, and the
VBLAST algorithm  proposed in \cite{Choi13}, as a function of energy per bit to noise power ratio
($E_{b}/N_{0}$) where $N_{0}$ is equal to $\sigma_w^2$.

ZF causes noise enhancement while it eliminates the ICI. Therefore,
it is seen that ZF performance is the worst. The reason for this can
also be explained by the ill-conditioned matrix $({\bf G}^{\dag}{\bf G})$ to be
inverted. This problem can be solved with MMSE, which provides a good
trade-off between ICI cancellation and noise elevation by using the
knowledge of the noise level. In other words, the noise enhancement
can be reduced by the insertion of the noise power in the inverse
matrix that given in Table 1.

We note that linear equalization of the received signal is
suboptimal as mentioned in Section II and hence Gibbs-based MAP
detection algorithm (MAP-GS) is proposed in Section III. It is
observed that MAP-GS outperforms both ZF and MMSE receivers while it
has similar performance to VBLAST. Moreover, the exact MAP
performance is also included to benchmark the proposed algorithms.
It can be concluded from these figures that the compared algorithms
have similar performance for the speed of 120 km/h while the performance
difference is obvious for the speed of 420 km/h. Moreover, it is seen that the
BER performance of all algorithms has slightly decreased for the BU
channel. In particular, it is observed that a savings in about 2 dB
is obtained at $BER=10^{-3}$, as compared with MMSE detection for the TU
channel.

It has been shown that when a proper detection technique is adopted, the
time-varying nature of the channel can be exploited as a provider of
time diversity\cite{Choi13}. In \cite{Choi13}, it was demonstrated
that VBLAST fully utilizes the time diversity while suppressing the
residual interference and the noise enhancement. Similarly to VBLAST,
we have seen that MAP-GS is also  a useful detection technique for
time-varying channels while having lower complexity. Therefore, in
particular, it is not surprising that in simulations the performance at
420km/h is better than that at 120km/h.

Finally, the BER performance of the proposed algorithm is presented
as a function of $Q$ in Fig. 4. The parameter $Q$ can be chosen to trade
off performance versus complexity. As a rule of thumb, we have seen that $Q=\lfloor f_{d_{max}}/\Delta f \rfloor +1$, where $f_{d_{max}}$ is the maximum
Doppler frequency and $\Delta f$ is the subcarrier spacing, is an
appropriate choice for Rayleigh fading\cite{schniter10}. In this
paper, $f_{d_{max}}/\Delta f$ values are given as the normalized
Doppler frequencies. It is concluded from these curves that the
selection of the $Q$ value is highly dependent on SNR values. In
particular, for $E_{b}/N_{0}=20 dB$, different $Q$ values show
similar performance because the effects of ICI are not very obvious
relative to the effects of the additive Gaussian noise. The $Q$ value has a greater role for $Eb/N_{0}$ above 25dB because ICI is
dominant. We note that $Q=3$ is sufficient for SNRs below 20dB.

\section{Conclusion}

Conventional detection methods such as ZF and MMSE have
irreducible error floors at high normalized Doppler frequency
$f_dT_s$ since ICI corrupts the orthogonality among subcarriers. On
the other hand, more sophisticated methods such as VBLAST require
too much complexity, especially for large numbers of
subcarriers. Therefore, we have proposed a new low-complexity
 maximum {\em a posteriori} probability (MAP) detection algorithm that
provides excellent performance with manageable complexity for OFDM
systems via the Gibbs sampling technique. In the simulation section, a
comparison with other previously known receiver structures has been made
and it has been demonstrated that MAP detection based on Gibbs sampling
provides performance that is close to that of the optimal MAP
detection algorithm for realistic fading conditions.


\bibliographystyle{IEEEtran}

%


\end{document}